\documentclass[english,aps,preprint,showpacs]{revtex4}
\usepackage[T1]{fontenc}
\usepackage[latin9]{inputenc}
\usepackage{color}
\usepackage{amsmath}
\usepackage{amssymb}
\usepackage{esint}

\makeatletter
\@ifundefined{textcolor}{}
{%
 \definecolor{BLACK}{gray}{0}
 \definecolor{WHITE}{gray}{1}
 \definecolor{RED}{rgb}{1,0,0}
 \definecolor{GREEN}{rgb}{0,1,0}
 \definecolor{BLUE}{rgb}{0,0,1}
 \definecolor{CYAN}{cmyk}{1,0,0,0}
 \definecolor{MAGENTA}{cmyk}{0,1,0,0}
 \definecolor{YELLOW}{cmyk}{0,0,1,0}
 }

\makeatother

\makeatother

\usepackage{babel}
\begin{document}

\title{The master equation for the reduced open-system dynamics, including
a Lindbladian description of finite-duration measurement}

\author{C. A. Brasil}

\email{carlosbrasil@ifsc.usp.br}

\selectlanguage{english}%

\author{R. d. J. Napolitano}

\affiliation{Instituto de Física de São Carlos, Universidade de São Paulo, P.O.
Box 369, 13560-970, São Carlos, SP, Brazil}
\begin{abstract}
We consider the problem of the measurement of a system occurring during
a finite time interval, while environmentally-induced noise decreases
the system-state coherence. We assume a Markovian measuring device
and, therefore, use a Lindbladian description for the measurement
dynamics. For studying the case of noise produced by a non-Markovian
environment, whose definition does not include the measuring apparatus,
we use the Redfield approach to the interaction between system and
environment. In the present hybrid theory, to trace out the environmental
degrees of freedom, we introduce an analytic method based on superoperator
algebra and Nakajima-Zwanzig projectors. The resulting master equation,
describing the reduced system dynamics, is illustrated in the case
of a qubit under phase noise during a finite-time measurement.
\end{abstract}


\maketitle

\section{Introduction}

Quantum dynamical semigroups \cite{key-1} constitute a powerful mathematical
formalism for the treatment of non-unitary processes in quantum mechanics,
which leads to many applications in the field of open systems, such
as analysis of dissipation, decoherence and quantum measurement theory
\cite{key-2}. It has been emphasized \cite{key-3,key-4} that such
phenomena can not have an adequate description by the conventional
formalism of quantum mechanics, based solely on the Liouville-von
Neumann equation.

The most general form for the generator of quantum dynamical semigroups
is the Lindblad\cite{key-5,key-6}, with several applications to Markovian
processes \cite{key-2} and, more recently, to non-Markovian situations
as well \cite{key-7,key-8}. Indeed, as demonstrated in \cite{key-2},
even in the case of the quantum Brownian motion it is possible to
transform the Caldeira-Leggett master equation \cite{key-9} into
a Lindblad equation, with the addition of a term that becomes small
in the high-temperature limit. Moreover, in its stochastic form, the
Lindblad equation can be used in numerical methods, such as in the
quantum-state diffusion approach \cite{key-10,key-11,key-12}.

Thus, the Lindblad equation is widely used to describe irreversible
quantum phenomena, such as the theoretical representation of finite-duration
measurements. If we assume that the environment of an open quantum
system under measurement does not include the measuring device, the
measurement can, therefore, be approached using a Lindblad equation
involving a set of Lindbladian operators taken as the measured observables,
but with the environmental degrees of freedom also included in the
global description. The next step to be taken in such a model is to
trace the environment variables out of the formulation, leaving only
a master equation describing both, the system and the set of measuring
observables. The tracing procedure is usually very complicated due
to the non-commutativity of several terms in the total Hamiltonian
comprising the system under scrutiny, its environment, and the observables
to be measured.

Moreover, in the finite-time measurement, we can consider a Markovian
interaction between the system to be measured and the measuring apparatus.
In that case, the relevant specification of the apparatus is effectively
contained in the Lindbladian and the problem can be treated with the
usual tools of that context. However, if we consider that the interaction
between the system and its environment is non-Markovian, the action
of the environmental degrees of freedon will be described by the Liouvillian
only.

In this case, the entanglement between the relevant system and its
environment complicates the description of the reduced time evolution
of the system. One way of taking into account the correlations between
the system and its environment involves the use of a projector superoperator
that transforms the total density operator $\hat{\rho}$ into its
reduced counterpart, $\hat{\rho}_{S}{,}$ calculated at the instant
under consideration, tensorially multiplied by the reduced density
matrix of the environment, $\hat{\rho}_{B}{,}$ calculated at the
initial time, that is, $\hat{\rho}\left(t\right)\mapsto\hat{\rho}_{S}\left(t\right)\otimes\hat{\rho}_{B}\left(0\right)$,
where the most popular approach is the one by Nakajima and Zwanzig
\cite{key-13,key-14}, who were motivated by the work of van Hove,
Prigogine, and Resibois \cite{key-15,key-16,key-17,key-18}. Even
so, however, it is still necessary to trace out the interaction to
obtain the temporal equation for the relevant part of the system.
Accomplish this task using directly the exponentials of each term
to suppress them makes the calculations extremely difficult and prohibitively
expensive. The procedure so hard, giving rise to many errors.

In this paper, we will analyze finite-time measurent for a case of
Markovian interaction between the principal system and the measurement
and non-Markovian interaction with the system and the environment.
We present a method to obtain the master equation in the Born-Markov
approximation, by tracing out the environmental degrees of freedom
using the time-independent thermodynamic projectors and superoperators
for each Hamiltonian (system, environment and interaction), including
the Lindbladian term. The method is easily demonstrated and the superoperator
properties substantially simplify the calculations, rendering the
resulting master equation compact and very simple.

Although we use the Born-Markov approximation and time-independent
thermodynamic projector superorerators, we believe it to be possible
to generalize the present approach to higher orders of the perturbation
series, as in \cite{key-19,key-20}, non-Markovian cases \cite{key-21,key-22,key-23,key-24},
or time-dependent thermodynamic projectors, as in\cite{key-25,key-26}.

The paper is structured as follows: the general problem is formulated
in the Sec. 2, with the \textcolor{red}{definition} of superoperators
and thermodynamic projectors; in Sec. 3, the necessary properties
of these superoperators are demonstrated; and, in Sec. 4, the master
equation is obtained. Finally, in Se. 5, we illustrate the method
in the case of a single qubit, subject to phase errors induced by
the environment, when a finite-duration measurement is performed on
an observable that commute with the unperturbed qubit Hamiltonian.

\section{Definitions}

The most general form for a master equation, according to the quantum
dynamical semigroups approach, is the Lindblad equation \cite{key-1,key-2,key-5,key-6},

\begin{multline}
\frac{d}{dt}\hat{\rho}_{SB}=-\frac{i}{\hbar}\left[\hat{H},\hat{\rho}_{SB}\right]+\underset{j}{\sum}\left(\hat{L}_{j}\hat{\rho}_{SB}\hat{L}_{j}^{\dagger}-\frac{1}{2}\left\{ \hat{L}_{j}^{\dagger}\hat{L}_{j},\hat{\rho}_{SB}\right\} \right),\label{lindblad}
\end{multline}
where $\hat{H}$ is the total Hamiltonian and the $\hat{L}_{j}$ are
the Lindblads (Hermitian operators for measurement description, not-Hermitian
for dissipation description) that represent a Markovian interaction.
Here we will consider the study of finite-time measurement. 

It is important to point out that the Lindbladian operator of Eq.
(\ref{lindblad}) appears as an effective device to emulate a measurement
on the system. Thus, the Lindbladian evolution plays the role of the
measuring-device action on the system. In fact, had we started from
an ab initio microscopic description of a measuring apparatus, we
would have to distinguish between its microscopic and macroscopic
variables. For each of the possible eigenvalues of the observable
being measured, there must correspond a single value of a macroscopic
variable. However, such a variable alone cannot completely describe
the state of the apparatus, for there is a large number of its microscopic
states that correspond to the same value of the macroscopic variable.
Since, in a measurement, the only variable that matters to the observer
is the macroscopic one, we can trace out the microscopic degrees of
freedom of the apparatus. This procedure results in a master equation
which, under the Born-Markov approximation \cite{key-2}, asssumes
the Lindblad form, which we adopt here \cite{key-27,key-28}. 

Let us consider that the Hamiltonian $\hat{H}$ can be written as

\[
\hat{H}=\hat{H}_{B}+\hat{H}_{SB}+\hat{H}_{S}{,}
\]
where $\hat{H}_{B}$ describes the environment, $\hat{H}_{S}$ describes
the relevant system, and $\hat{H}_{SB}$ is the Hamiltonian for the
system-environment non-Markovian interaction. Here we assume, as usual,
that the environment-system interaction is factorizable, i.e.,
\begin{eqnarray}
\hat{H}_{SB} & = & \sum_{k}\hat{S}_{k}\hat{B}_{k},\label{defHSB}
\end{eqnarray}
 where, for each $k$, $\hat{S}_{k}$ operates only on the system
and $\hat{B}_{k}$, only on the environment. The form of the interaction,
Eq. (\ref{defHSB}), is general, satisfied by both \emph{amplitude}
and \emph{phase damping models}. By hypothesis, the Lindblads $\hat{L}_{j}$
will operate only on the system $S$. Our aim is to obtain an equation
for the time evolution of the system reduced density matrix, $\hat{\rho}_{S}$,

\[
\hat{\rho}_{S}\left(t\right)=tr_{B}\left\{ \hat{\rho}_{SB}\left(t\right)\right\} {.}
\]

Let us begin by defining some superoperators. For any density-matrix
operator $\hat{X}$,
\begin{eqnarray}
\mathcal{B}\hat{X} & = & -\frac{i}{\hbar}\left[\hat{H}_{B},\hat{X}\right]{,}\label{defB}
\end{eqnarray}
 
\begin{eqnarray}
\mathcal{S}\hat{X} & = & -\frac{i}{\hbar}\left[\hat{H}_{S},\hat{X}\right]+\underset{j}{\sum}\left(\hat{L}_{j}\hat{X}\hat{L}_{j}^{\dagger}-\frac{1}{2}\left\{ \hat{L}_{j}^{\dagger}\hat{L}_{j},\hat{X}\right\} \right){,}\label{defS}
\end{eqnarray}
 and
\begin{eqnarray}
\mathcal{F}\hat{X} & = & -\frac{i}{\hbar}\left[\hat{H}_{SB},\hat{X}\right]{.}\label{defF}
\end{eqnarray}

We will also use the Nakajima-Zwanzig thermodynamic projectors $\mathcal{P}$
and $\mathcal{Q}$ \cite{key-2,key-13,key-14}. $\mathcal{{P}}$ is
such that its action is defined, for any density operator ${{\hat{{\rho}}}_{B}\left(t_{0}\right),}$
by 
\begin{eqnarray}
\mathcal{P}\hat{X}\left(t\right) & = & \hat{\rho}_{B}\left(t_{0}\right)\otimes\mathrm{Tr}_{B}\left\{ \hat{X}\left(t\right)\right\} {.}\label{defP}
\end{eqnarray}
 It is easy to check that
\begin{eqnarray*}
\mathcal{P}^{2} & = & \mathcal{P}{.}
\end{eqnarray*}
 We also define
\begin{eqnarray}
\mathcal{Q} & = & \mathcal{I}-\mathcal{P}{,}\label{defQ}
\end{eqnarray}
 where $\mathcal{I}$ is the identity superoperator ($\mathcal{I}\hat{X}\left(t\right)=\hat{X}\left(t\right)$).
It follows from that definition that $\mathcal{{Q}}$ is also a projector,
i.e.,
\begin{eqnarray*}
\mathcal{Q}^{2} & = & \mathcal{Q}.
\end{eqnarray*}

\section{The superoperator properties of $\mathcal{{B}}$, $\mathcal{{S}}$
and $\mathcal{{F}}$ }

As a consequence of the definitions (\ref{defB}) and (\ref{defS}),
$\mathcal{B}$ acts only on the environment and $\mathcal{S}{,}$
only on the system. Hence, 
\begin{eqnarray*}
\mathcal{B}\mathcal{S} & = & \mathcal{S}\mathcal{B}.
\end{eqnarray*}
 It follows from that commutation relation that 
\begin{eqnarray*}
\exp\left(\mathcal{S}t+\mathcal{B}t\right) & = & \exp\left(\mathcal{S}t\right)\exp\left(\mathcal{B}t\right)=\exp\left(\mathcal{B}t\right)\exp\left(\mathcal{S}t\right){.}
\end{eqnarray*}

For simplicity, let us consider the initial time as \emph{zero} ($t_{0}=0$)
and that the global density operator is initially factorized:
\begin{eqnarray*}
\hat{\rho}_{SB}\left(0\right) & = & \hat{\rho}_{S}\left(0\right)\otimes\hat{\rho}_{B}\left(0\right){.}
\end{eqnarray*}
 For the sake of simplicity, let us ignore the $\otimes$ symbol henceforth
and write
\begin{eqnarray*}
\hat{\rho}_{SB}\left(0\right) & = & \hat{\rho}_{S}\left(0\right)\hat{\rho}_{B}\left(0\right)=\hat{\rho}_{B}\left(0\right)\hat{\rho}_{S}\left(0\right).
\end{eqnarray*}

The partial trace, over the environmental degrees of freedom, of the
resulting action of $\exp\left(-\mathcal{B}t\right)$ on the global
density operator, can be written as
\begin{eqnarray*}
\mathrm{Tr}_{B}\left\{ \exp\left(-\mathcal{B}t\right)\hat{\rho}_{SB}\left(t\right)\right\}  & = & \sum_{n=0}^{\infty}\frac{\left(-t\right)^{n}}{n!}\mathrm{Tr}_{B}\left\{ \mathcal{B}^{n}\hat{\rho}_{SB}\left(t\right)\right\} \\
 & = & \mathrm{Tr}_{B}\left\{ \hat{\rho}_{SB}\left(t\right)\right\} +\sum_{n=1}^{\infty}\frac{\left(-t\right)^{n}}{n!}\mathrm{Tr}_{B}\left\{ \mathcal{B}^{n}\hat{\rho}_{SB}\left(t\right)\right\} .
\end{eqnarray*}
 But,
\begin{eqnarray}
\mathrm{Tr}_{B}\left\{ \mathcal{B}^{n}\hat{\rho}_{SB}\left(t\right)\right\}  & = & \mathrm{Tr}_{B}\left\{ \mathcal{B}\mathcal{B}^{n-1}\hat{\rho}_{SB}\left(t\right)\right\} \nonumber \\
 & = & -\frac{i}{\hbar}\mathrm{Tr}_{B}\left\{ \left[\hat{H}_{B},\mathcal{B}^{n-1}\hat{\rho}_{SB}\left(t\right)\right]\right\} {.}\label{aux1}
\end{eqnarray}
 The trace is basis independent and, therefore, it is convenient,
for the treatment of Eq. (\ref{aux1}), to use the environmental basis
$\left\{ \left|k\right\rangle \right\} $ of eigenstates of $\hat{H}_{B}$,
i.e.,
\begin{eqnarray*}
\hat{H}_{B}\left|k\right\rangle  & = & E_{k}^{B}\left|k\right\rangle ,
\end{eqnarray*}
 for all $k$ (where the eingenstats can be degenerate or not). With
this consideration,
\begin{eqnarray*}
\mathrm{Tr}_{B}\left\{ \mathcal{B}^{n}\hat{\rho}_{SB}\left(t\right)\right\}  & = & -\frac{i}{\hbar}\sum_{k}\left\langle k\right|\left[\hat{H}_{B},\mathcal{B}^{n-1}\hat{\rho}_{SB}\left(t\right)\right]\left|k\right\rangle \\
 & = & -\frac{i}{\hbar}\left\langle k\right|\hat{H}_{B}\mathcal{B}^{n-1}\hat{\rho}_{SB}\left(t\right)\left|k\right\rangle -\left\langle k\right|\mathcal{B}^{n-1}\hat{\rho}_{SB}\left(t\right)\hat{H}_{B}\left|k\right\rangle \\
 & = & -\frac{i}{\hbar}\left\langle k\right|E_{k}^{B}\mathcal{B}^{n-1}\hat{\rho}_{SB}\left(t\right)\left|k\right\rangle -\left\langle k\right|\mathcal{B}^{n-1}\hat{\rho}_{SB}\left(t\right)E_{k}^{B}\left|k\right\rangle \\
 & = & -\frac{i}{\hbar}\left[E_{k}^{B}\left\langle k\right|\mathcal{B}^{n-1}\hat{\rho}_{SB}\left(t\right)\left|k\right\rangle -E_{k}^{B}\left\langle k\right|\mathcal{B}^{n-1}\hat{\rho}_{SB}\left(t\right)\left|k\right\rangle \right]\\
 & = & 0{,}\textrm{ for {all} }k.
\end{eqnarray*}
 Therefore,
\begin{eqnarray}
\mathrm{Tr}_{{\normalcolor {\normalcolor B}}}\left\{ \exp\left(-\mathcal{B}t\right)\hat{\rho}_{SB}\left(t\right)\right\}  & = & \mathrm{Tr}_{B}\left\{ \hat{\rho}_{SB}\left(t\right)\right\} +\sum_{n=1}^{\infty}\frac{\left(-t\right)^{n}}{n!}\mathrm{Tr}_{B}\left\{ \mathcal{B}^{n}\hat{\rho}_{SB}\left(t\right)\right\} \nonumber \\
 & = & \mathrm{Tr}_{B}\left\{ \hat{\rho}_{SB}\left(t\right)\right\} +0\nonumber \\
 & = & \mathrm{Tr}_{B}\left\{ \hat{\rho}_{SB}\left(t\right)\right\} .\label{prop1}
\end{eqnarray}
 From this property, it follows that
\begin{eqnarray}
\exp\left(-\mathcal{S}t\right)\hat{\rho}_{S}\left(t\right) & = & \exp\left(-\mathcal{S}t\right)\mathrm{Tr}_{B}\left\{ \hat{\rho}_{SB}\left(t\right)\right\} \nonumber \\
 & = & \exp\left(-\mathcal{S}t\right)\mathrm{Tr}_{B}\left\{ \exp\left(-\mathcal{B}t\right)\hat{\rho}_{SB}\left(t\right)\right\} \nonumber \\
 & = & \mathrm{Tr}_{B}\left\{ \exp\left(-\mathcal{S}t\right)\exp\left(-\mathcal{B}t\right)\hat{\rho}_{SB}\left(t\right)\right\} \nonumber \\
 & = & \mathrm{Tr}_{B}\left\{ \exp\left(-\mathcal{S}t-\mathcal{B}t\right)\hat{\rho}_{SB}\left(t\right)\right\} .\label{prop2}
\end{eqnarray}

For the interaction superoperator of Eq. (\ref{defF}), we have, from
the (\ref{defHSB}) hypothesis,
\begin{eqnarray*}
\mathcal{F}\hat{X} & = & -\frac{i}{\hbar}\left[\sum_{k}\hat{S}_{k}\hat{B}_{k},\hat{X}\right]\\
 & = & -\frac{i}{\hbar}\sum_{k}\left[\hat{S}_{k}\hat{B}_{k},\hat{X}\right]
\end{eqnarray*}
 for any density operator $\hat{X}$.

\section{Tracing out the environmental degrees of freedom}

The equation that we want to solve is
\begin{eqnarray*}
\frac{d}{dt}\hat{\rho}_{SB}\left(t\right) & = & -\frac{i}{\hbar}\left[\hat{H}_{S}+\hat{H}_{B}+\hat{H}_{SB},\hat{\rho}_{SB}\left(t\right)\right]+\underset{j}{\sum}\left(\hat{L}_{j}\hat{\rho}_{SB}\hat{L}_{j}^{\dagger}-\frac{1}{2}\left\{ \hat{L}_{j}^{\dagger}\hat{L}_{j},\hat{\rho}_{SB}\right\} \right){.}
\end{eqnarray*}
 In terms of the superoperators of Eqs. (\ref{defB}), (\ref{defS}),
and (\ref{defF}), the equation becomes
\begin{eqnarray}
\frac{d}{dt}\hat{\rho}_{SB}\left(t\right) & = & \left(\mathcal{S}+\mathcal{B}+\mathcal{F}\right)\hat{\rho}_{SB}\left(t\right).\label{eqsuper}
\end{eqnarray}
 With the properties expressed by Eqs. (\ref{prop1}) and (\ref{prop2}),
let us define the operator
\begin{eqnarray}
\hat{\alpha}\left(t\right) & = & \exp\left(-\mathcal{S}t-\mathcal{B}t\right)\hat{\rho}_{SB}\left(t\right).\label{defalfa}
\end{eqnarray}
 Then,
\begin{eqnarray}
\hat{\rho}_{S}\left(t\right) & = & \exp\left(\mathcal{S}t\right)\mathrm{Tr}_{B}\left\{ \hat{\alpha}\left(t\right)\right\} \label{invalfa}
\end{eqnarray}
 and the equation satisfied by $\hat{\alpha}\left(t\right)$ is written
as 
\begin{eqnarray}
\frac{d}{dt}\hat{\alpha}\left(t\right) & = & \frac{d}{dt}\left[\exp\left(-\mathcal{S}t-\mathcal{B}t\right)\hat{\rho}_{SB}\left(t\right)\right]\nonumber \\
 & = & \left[\frac{d}{dt}\exp\left(-\mathcal{S}t-\mathcal{B}t\right)\right]\hat{\rho}_{SB}\left(t\right)+\exp\left(-\mathcal{S}t-\mathcal{B}t\right)\frac{d}{dt}\hat{\rho}_{SB}\left(t\right)\nonumber \\
 & = & -\left(\mathcal{S}+B\right)\exp\left(-\mathcal{S}t-Bt\right)\hat{\rho}_{SB}\left(t\right)+\exp\left(-\mathcal{S}t-\mathcal{B}t\right)\frac{d}{dt}\hat{\rho}_{SB}\left(t\right).\label{aux2}
\end{eqnarray}

We can insert Eq. (\ref{eqsuper}) into Eq. (\ref{aux2}) and the
result is
\begin{eqnarray*}
\frac{d}{dt}\hat{\alpha}\left(t\right) & = & \exp\left(-\mathcal{S}t-\mathcal{B}t\right)\mathcal{F}\exp\left(\mathcal{S}t+\mathcal{B}t\right)\hat{\alpha}\left(t\right).
\end{eqnarray*}
 Let us define, then, the superoperator:
\begin{eqnarray}
\mathcal{G}\left(t\right) & = & \exp\left(-\mathcal{S}t-\mathcal{B}t\right)\mathcal{F}\exp\left(\mathcal{S}t+\mathcal{B}t\right).\label{defG}
\end{eqnarray}
 We note that $\mathcal{G}\left(t\right)$ is of the order of magnitude
of the interaction $\hat{H}_{SB}$, because it is proportional to
the super-operator $\mathcal{F}$ that, in turn, is of the order of
magnitude of ${\hat{H}}_{{SB}}{.}$ Thus, we have
\begin{eqnarray}
\frac{d}{dt}\hat{\alpha}\left(t\right) & = & \mathcal{G}\left(t\right)\hat{\alpha}\left(t\right).\label{eqalfa2}
\end{eqnarray}

Up to this moment, we have used superoperators based on each of the
Hamiltonian and Lindbladian terms ($\mathcal{B}$, $\mathcal{S}$
e $\mathcal{F}$) and obtained Eq. (\ref{eqalfa2}), which involves
the global density operator $\hat{\rho}_{SB}\left(t\right)$. Now,
let us use the Nakajima-Zwanzig thermodynamic projectors, defined
by Eqs. (\ref{defP}) and (\ref{defQ}), in Eq. (\ref{eqalfa2}):
\begin{eqnarray*}
\mathcal{P}\frac{d}{dt}\hat{\alpha}\left(t\right) & = & \mathcal{P}\mathcal{G}\left(t\right)\hat{\alpha}\left(t\right)\\
 & = & \mathcal{P}\mathcal{G}\left(t\right)\left(\mathcal{P}+\mathcal{Q}\right)\hat{\alpha}\left(t\right)\\
 & = & \mathcal{P}\mathcal{G}\left(t\right)\mathcal{P}\hat{\alpha}\left(t\right)+\mathcal{P}\mathcal{G}\left(t\right)\mathcal{Q}\hat{\alpha}\left(t\right)\\
 & = & \mathcal{P}\mathcal{G}\left(t\right)\mathcal{P}^{2}\hat{\alpha}\left(t\right)+\mathcal{P}\mathcal{G}\left(t\right)\mathcal{Q}^{2}\hat{\alpha}\left(t\right)
\end{eqnarray*}
 and
\begin{eqnarray*}
\mathcal{Q}\frac{d}{dt}\hat{\alpha}\left(t\right) & = & \mathcal{Q}\mathcal{G}\left(t\right)\hat{\alpha}\left(t\right)\\
 & = & \mathcal{Q}\mathcal{G}\left(t\right)\left(\mathcal{P}+\mathcal{Q}\right)\hat{\alpha}\left(t\right)\\
 & = & \mathcal{Q}\mathcal{G}\left(t\right)\mathcal{P}\hat{\alpha}\left(t\right)+\mathcal{Q}\mathcal{G}\left(t\right)\mathcal{Q}\hat{\alpha}\left(t\right)\\
 & = & \mathcal{Q}\mathcal{G}\left(t\right)\mathcal{P}^{2}\hat{\alpha}\left(t\right)+\mathcal{Q}\mathcal{G}\left(t\right)\mathcal{Q}^{2}\hat{\alpha}\left(t\right).
\end{eqnarray*}
 Since those projectors are time-independent, it follows that

\begin{equation}
\begin{cases}
\frac{d}{dt}\left[\mathcal{P}\alpha\left(t\right)\right]= & \left[\mathcal{P}\mathcal{G}\left(t\right)\mathcal{P}\right]\left[\mathcal{P}\alpha\left(t\right)\right]+\left[\mathcal{P}\mathcal{G}\left(t\right)\mathcal{Q}\right]\left[\mathcal{Q}\alpha\left(t\right)\right]{,}\\
\frac{d}{dt}\left[\mathcal{Q}\alpha\left(t\right)\right]= & \left[\mathcal{Q}\mathcal{G}\left(t\right)\mathcal{P}\right]\left[\mathcal{P}\alpha\left(t\right)\right]+\left[\mathcal{Q}\mathcal{G}\left(t\right)\mathcal{Q}\right]\left[\mathcal{Q}\alpha\left(t\right)\right]{.}
\end{cases}\label{sist1}
\end{equation}
 Let us formally integrate the second of the Eqs. (\ref{sist1}):
\begin{eqnarray*}
\mathcal{Q}\hat{\alpha}\left(t\right) & = & \mathcal{Q}\hat{\alpha}\left(0\right)+\int_{0}^{t}dt^{\prime}\,\left[\mathcal{Q}\mathcal{G}\left(t^{\prime}\right)\mathcal{P}\right]\left[\mathcal{P}\hat{\alpha}\left(t^{\prime}\right)\right]+\int_{0}^{t}dt^{\prime}\,\left[\mathcal{Q}\mathcal{G}\left(t^{\prime}\right)\mathcal{Q}\right]\left[\mathcal{Q}\hat{\alpha}\left(t^{\prime}\right)\right].
\end{eqnarray*}
 From the definition of $\hat{\alpha}\left(t\right){,}$ Eq. (\ref{defalfa}),
we know that
\begin{eqnarray*}
\hat{\alpha}\left(0\right) & = & \hat{\rho}_{SB}\left(0\right)\\
 & = & \hat{\rho}_{S}\left(0\right)\hat{\rho}_{B}\left(0\right)
\end{eqnarray*}
 and, therefore,
\begin{eqnarray*}
\mathcal{Q}\hat{\alpha}\left(0\right) & = & \left(\mathcal{I}-\mathcal{P}\right)\hat{\rho}_{S}\left(0\right)\hat{\rho}_{B}\left(0\right)\\
 & = & \hat{\rho}_{S}\left(0\right)\hat{\rho}_{B}\left(0\right)-\mathcal{P}\hat{\rho}_{S}\left(0\right)\hat{\rho}_{B}\left(0\right)\\
 & = & \hat{\rho}_{S}\left(0\right)\hat{\rho}_{B}\left(0\right)-\hat{\rho}_{B}\left(0\right)\mathrm{Tr}_{B}\left\{ \hat{\rho}_{S}\left(0\right)\hat{\rho}_{B}\left(0\right)\right\} \\
 & = & \hat{\rho}_{S}\left(0\right)\hat{\rho}_{B}\left(0\right)-\hat{\rho}_{B}\left(0\right)\hat{\rho}_{S}\left(0\right)\mathrm{Tr}_{B}\left\{ \hat{\rho}_{B}\left(0\right)\right\} \\
 & = & \hat{\rho}_{S}\left(0\right)\hat{\rho}_{B}\left(0\right)-\hat{\rho}_{B}\left(0\right)\hat{\rho}_{S}\left(0\right)\\
 & = & 0.
\end{eqnarray*}
 Hence,
\begin{eqnarray}
\mathcal{Q}\hat{\alpha}\left(t\right) & = & \int_{0}^{t}dt^{\prime}\,\left[\mathcal{Q}\mathcal{G}\left(t^{\prime}\right)\mathcal{P}\right]\left[\mathcal{P}\hat{\alpha}\left(t^{\prime}\right)\right]+\int_{0}^{t}dt^{\prime}\,\left[\mathcal{Q}\mathcal{G}\left(t^{\prime}\right)\mathcal{Q}\right]\left[\mathcal{Q}\hat{\alpha}\left(t^{\prime}\right)\right],\label{eqQalfa1}
\end{eqnarray}
 showing that $\mathcal{Q}\hat{\alpha}\left(t\right)$ is of the order
of magnitude of ${\hat{H}}_{{S{\normalcolor {\normalcolor B}}}}{.}$
Here we use the Born approximation and only keep terms up to the second
order of ${\hat{H}}_{{SB}}{.}$ Accordingly, the second-order iteration
of Eq. (\ref{eqQalfa1}) gives
\begin{eqnarray}
\mathcal{Q}\hat{\alpha}\left(t\right) & = & \int_{0}^{t}dt^{\prime}\,\left[\mathcal{Q}\mathcal{G}\left(t^{\prime}\right)\mathcal{P}\right]\left[\mathcal{P}\hat{\alpha}\left(t^{\prime}\right)\right]\nonumber \\
 & + & \int_{0}^{t}dt^{\prime}\,\left[\mathcal{Q}\mathcal{G}\left(t^{\prime}\right)\mathcal{Q}\right]\left[\int_{0}^{t^{\prime}}dt^{\prime\prime}\,\left[\mathcal{Q}\mathcal{G}\left(t^{\prime\prime}\right)\mathcal{P}\right]\left[\mathcal{P}\hat{\alpha}\left(t^{\prime\prime}\right)\right]\right].\label{eqQalfa2}
\end{eqnarray}

Let us formally integrate the first of the Eqs. (\ref{sist1}):
\begin{eqnarray}
\mathcal{P}\hat{\alpha}\left(t\right) & = & \mathcal{P}\hat{\alpha}\left(0\right)+\int_{0}^{t}dt^{\prime}\,\left[\mathcal{P}\mathcal{G}\left(t^{\prime}\right)\mathcal{P}\right]\left[\mathcal{P}\hat{\alpha}\left(t^{\prime}\right)\right]+\int_{0}^{t}dt^{\prime}\,\left[\mathcal{P}\mathcal{G}\left(t^{\prime}\right)\mathcal{Q}\right]\left[\mathcal{Q}\hat{\alpha}\left(t^{\prime}\right)\right].\label{eqPalfa1}
\end{eqnarray}
 Substituing Eq. (\ref{eqPalfa1}) in the first of the Eqs. (\ref{sist1}),
we obtain
\begin{eqnarray}
\frac{d}{dt}\left[\mathcal{P}\hat{\alpha}\left(t\right)\right] & = & \left[\mathcal{P}\mathcal{G}\left(t\right)\mathcal{P}\right]\left[\mathcal{P}\hat{\alpha}\left(0\right)\right]+\left[\mathcal{P}\mathcal{G}\left(t\right)\mathcal{P}\right]\int_{0}^{t}dt^{\prime}\,\left[\mathcal{P}\mathcal{G}\left(t^{\prime}\right)\mathcal{P}\right]\left[\mathcal{P}\hat{\alpha}\left(t^{\prime}\right)\right]\nonumber \\
 & + & \left[\mathcal{P}\mathcal{G}\left(t\right)\mathcal{P}\right]\int_{0}^{t}dt^{\prime}\,\left[\mathcal{P}\mathcal{G}\left(t^{\prime}\right)\mathcal{Q}\right]\left[\mathcal{Q}\hat{\alpha}\left(t^{\prime}\right)\right]+\left[\mathcal{P}\mathcal{G}\left(t\right)\mathcal{Q}\right]\left[\mathcal{Q}\hat{\alpha}\left(t\right)\right].\label{eqPalfa2}
\end{eqnarray}
 From Eq. (\ref{eqQalfa2}) we see that the third term on the right-hand
side of Eq. (\ref{eqPalfa2}) is of the third order in ${\hat{H}}_{{S{\normalcolor {\normalcolor {\normalcolor B}}}}}{,}$
and, therefore, we neglect it. Substituing Eq. (\ref{eqQalfa2}) in
the forth term of right-hand side of Eq. (\ref{eqPalfa2}) and keeping
only contributions up to second order in ${\hat{H}}_{{S{\normalcolor {\normalcolor {\normalcolor B}}}}}{,}$
we obtain
\begin{eqnarray*}
\frac{d}{dt}\left[\mathcal{P}\hat{\alpha}\left(t\right)\right] & = & \left[\mathcal{P}\mathcal{G}\left(t\right)\mathcal{P}\right]\left[\mathcal{P}\hat{\alpha}\left(0\right)\right]+\left[\mathcal{P}\mathcal{G}\left(t\right)\mathcal{P}\right]\int_{0}^{t}dt^{\prime}\,\left[\mathcal{P}\mathcal{G}\left(t^{\prime}\right)\mathcal{P}\right]\left[\mathcal{P}\alpha\hat{}\left(t^{\prime}\right)\right]\\
 & + & \left[\mathcal{P}\mathcal{G}\left(t\right)\mathcal{Q}\right]\int_{0}^{t}dt^{\prime}\,\left[\mathcal{Q}\mathcal{G}\left(t^{\prime}\right)\mathcal{P}\right]\left[\mathcal{P}\hat{\alpha}\left(t^{\prime}\right)\right]\\
 & = & \left[\mathcal{P}\mathcal{G}\left(t\right)\mathcal{P}\right]\left[\mathcal{P}\hat{\alpha}\left(0\right)\right]+\int_{0}^{t}dt^{\prime}\,\left[\mathcal{P}\mathcal{G}\left(t\right)\mathcal{P}\right]\left[\mathcal{P}\mathcal{G}\left(t^{\prime}\right)\mathcal{P}\right]\left[\mathcal{P}\hat{\alpha}\left(t^{\prime}\right)\right]\\
 & + & \int_{0}^{t}dt^{\prime}\,\left[\mathcal{P}\mathcal{G}\left(t\right)\mathcal{Q}\right]\left[\mathcal{Q}\mathcal{G}\left(t^{\prime}\right)\mathcal{P}\right]\left[\mathcal{P}\hat{\alpha}\left(t^{\prime}\right)\right]\\
 & = & \left[\mathcal{P}\mathcal{G}\left(t\right)\mathcal{P}\right]\left[\hat{\mathcal{P}\alpha}\left(0\right)\right]+\int_{0}^{t}dt^{\prime}\,\left[\mathcal{P}\mathcal{G}\left(t\right)\mathcal{P}\mathcal{P}\mathcal{G}\left(t^{\prime}\right)\mathcal{P}\right]\left[\mathcal{P}\hat{\alpha}\left(t^{\prime}\right)\right]\\
 & + & \int_{0}^{t}dt^{\prime}\,\left[\mathcal{P}\mathcal{G}\left(t\right)\mathcal{Q}\mathcal{Q}\mathcal{G}\left(t^{\prime}\right)\mathcal{P}\right]\left[\mathcal{P}\hat{\alpha}\left(t^{\prime}\right)\right]\\
 & = & \left[\mathcal{P}\mathcal{G}\left(t\right)\mathcal{P}\right]\left[\mathcal{P}\hat{\alpha}\left(0\right)\right]+\int_{0}^{t}dt^{\prime}\,\left[\mathcal{P}\mathcal{G}\left(t\right)\mathcal{P}\mathcal{G}\left(t^{\prime}\right)\mathcal{P}\right]\left[\mathcal{P}\hat{\alpha}\left(t^{\prime}\right)\right]\\
 & + & \int_{0}^{t}dt^{\prime}\,\left[\mathcal{P}\mathcal{G}\left(t\right)\mathcal{Q}\mathcal{G}\left(t^{\prime}\right)\mathcal{P}\right]\left[\mathcal{P}\hat{\alpha}\left(t^{\prime}\right)\right]\\
 & = & \left[\mathcal{P}\mathcal{G}\left(t\right)\mathcal{P}\right]\left[\mathcal{P}\hat{\alpha}\left(0\right)\right]\\
 & + & \int_{0}^{t}dt^{\prime}\,\left[\mathcal{P}\mathcal{G}\left(t\right)\mathcal{P}\mathcal{G}\left(t^{\prime}\right)\mathcal{P}+\mathcal{P}\mathcal{G}\left(t\right)\mathcal{Q}\mathcal{G}\left(t^{\prime}\right)\mathcal{P}\right]\left[\mathcal{P}\hat{\alpha}\left(t^{\prime}\right)\right]\\
 & = & \left[\mathcal{P}\mathcal{G}\left(t\right)\mathcal{P}\right]\left[\mathcal{P}\hat{\alpha}\left(0\right)\right]\\
 & + & \int_{0}^{t}dt^{\prime}\,\left[\mathcal{P}\mathcal{G}\left(t\right)\left(\mathcal{P}+\mathcal{Q}\right)\mathcal{G}\left(t^{\prime}\right)\mathcal{P}\right]\left[\mathcal{P}\hat{\alpha}\left(t^{\prime}\right)\right],
\end{eqnarray*}
 i.e.,
\begin{eqnarray*}
\frac{d}{dt}\left[\mathcal{P}\hat{\alpha}\left(t\right)\right] & = & \left[\mathcal{P}\mathcal{G}\left(t\right)\mathcal{P}\right]\left[\mathcal{P}\hat{\alpha}\left(0\right)\right]+\int_{0}^{t}dt^{\prime}\,\left[\mathcal{P}\mathcal{G}\left(t\right)\mathcal{G}\left(t^{\prime}\right)\mathcal{P}\right]\left[\mathcal{P}\hat{\alpha}\left(t^{\prime}\right)\right].
\end{eqnarray*}
 The first term of this equation can be written as 
\begin{eqnarray*}
\left[\mathcal{P}\mathcal{G}\left(t\right)\mathcal{P}\right]\left[\mathcal{P}\hat{\alpha}\left(0\right)\right] & = & \mathcal{P}\mathcal{G}\left(t\right)\mathcal{P}\mathcal{P}\hat{\alpha}\left(0\right)\\
 & = & \mathcal{P}\mathcal{G}\left(t\right)\mathcal{P}\hat{\alpha}\left(0\right)\\
 & = & \mathcal{P}\mathcal{G}\left(t\right)\mathcal{P}\hat{\rho}_{S}\left(0\right)\hat{\rho}_{B}\left(0\right)\\
 & = & \mathcal{P}\mathcal{G}\left(t\right)\hat{\rho}_{B}\left(0\right)\mathrm{Tr}_{B}\left\{ \hat{\rho}_{S}\left(0\right)\hat{\rho}_{B}\left(0\right)\right\} \\
 & = & \mathcal{P}\mathcal{G}\left(t\right)\hat{\rho}_{B}\left(0\right)\hat{\rho}_{S}\left(0\right)\mathrm{Tr}_{B}\left\{ \hat{\rho}_{B}\left(0\right)\right\} \\
 & = & \mathcal{P}\mathcal{G}\left(t\right)\hat{\rho}_{B}\left(0\right)\hat{\rho}_{S}\left(0\right)\\
 & = & \hat{\rho}_{B}\left(0\right)\mathrm{Tr}_{B}\left\{ \mathcal{G}\left(t\right)\hat{\rho}_{B}\left(0\right)\hat{\rho}_{S}\left(0\right)\right\} \\
 & = & \hat{\rho}_{B}\left(0\right)\mathrm{Tr}_{B}\left\{ \mathcal{G}\left(t\right)\hat{\rho}_{B}\left(0\right)\right\} \hat{\rho}_{S}\left(0\right).
\end{eqnarray*}
 But, following from Eq. (\ref{defG}),
\begin{eqnarray*}
\mathrm{Tr}_{B}\left\{ \mathcal{G}\left(t\right)\hat{\rho}_{B}\left(0\right)\right\} \hat{\rho}_{S}\left(0\right) & = & \mathrm{Tr}_{B}\left\{ \exp\left(-\mathcal{S}t-\mathcal{B}t\right)\mathcal{F}\exp\left(\mathcal{S}t+\mathcal{B}t\right)\hat{\rho}_{B}\left(0\right)\right\} \hat{\rho}_{S}\left(0\right)\\
 & = & \mathrm{Tr}_{B}\left\{ \exp\left(-\mathcal{S}t\right)\exp\left(-\mathcal{B}t\right)\mathcal{F}\exp\left(Bt\right)\exp\left(\mathcal{S}t\right)\hat{\rho}_{B}\left(0\right)\right\} \hat{\rho}_{S}\left(0\right)\\
 & = & \exp\left(-\mathcal{S}t\right)\mathrm{Tr}_{B}\left\{ \exp\left(-\mathcal{B}t\right)\mathcal{F}\exp\left(\mathcal{B}t\right)\hat{\rho}_{B}\left(0\right)\right\} \exp\left(\mathcal{S}t\right)\hat{\rho}_{S}\left(0\right){,}
\end{eqnarray*}
 i.e.,
\begin{eqnarray*}
\left[\mathcal{P}\mathcal{G}\left(t\right)\mathcal{P}\right]\left[\mathcal{P}\hat{\alpha}\left(0\right)\right] & = & \hat{\rho}_{B}\left(0\right)\exp\left(-\mathcal{S}t\right)\mathrm{Tr}_{B}\left\{ \exp\left(-\mathcal{B}t\right)\mathcal{F}\exp\left(\mathcal{B}t\right)\hat{\rho}_{B}\left(0\right)\right\} \exp\left(\mathcal{S}t\right)\hat{\rho}_{S}\left(0\right).
\end{eqnarray*}

Now, let us use Eq. (\ref{defHSB}):
\begin{eqnarray*}
\mathrm{Tr}_{B}\left\{ \exp\left(-\mathcal{B}t\right)\mathcal{F}\exp\left(\mathcal{B}t\right)\hat{\rho}_{B}\left(0\right)\right\}  & = & -\frac{i}{\hbar}\sum_{k}\mathrm{Tr}_{B}\left\{ \exp\left(-\mathcal{B}t\right)\left[S_{k}B_{k},\left[\exp\left(\mathcal{B}t\right)\hat{\rho}_{B}\left(0\right)\right]\right]\right\} \\
 & = & -\frac{i}{\hbar}\sum_{k}\mathrm{Tr}_{B}\left\{ \exp\left(-\mathcal{B}t\right)S_{k}\left[B_{k},\left[\exp\left(\mathcal{B}t\right)\hat{\rho}_{B}\left(0\right)\right]\right]\right\} \\
 & = & -\frac{i}{\hbar}\sum_{k}S_{k}\mathrm{Tr}_{B}\left\{ \exp\left(-\mathcal{B}t\right)\left[B_{k},\left[\exp\left(\mathcal{B}t\right)\hat{\rho}_{B}\left(0\right)\right]\right]\right\} .
\end{eqnarray*}
 Analogously to the calculation leading to Eq. (\ref{prop1}), we
obtain
\begin{eqnarray*}
\mathrm{Tr}_{B}\left\{ \exp\left(-\mathcal{B}t\right)\left[\hat{B}_{k},\left[\exp\left(\mathcal{B}t\right)\hat{\rho}_{B}\left(0\right)\right]\right]\right\}  & = & \mathrm{Tr}_{B}\left\{ \left[\hat{B}_{k},\left[\exp\left(\mathcal{B}t\right)\hat{\rho}_{B}\left(0\right)\right]\right]\right\} \\
 & = & \mathrm{Tr}_{B}\left\{ \hat{B}_{k}\left[\exp\left(\mathcal{B}t\right)\hat{\rho}_{B}\left(0\right)\right]\right\} \\
 & - & \mathrm{Tr}_{B}\left\{ \left[\exp\left(\mathcal{B}t\right)\hat{\rho}_{B}\left(0\right)\right]\hat{B}_{k}\right\} \\
 & = & 0{,}
\end{eqnarray*}
 which implies that

\[
\left[\mathcal{P}\mathcal{G}\left(t\right)\mathcal{P}\right]\left[\mathcal{P}\hat{\alpha}\left(0\right)\right]=0
\]
 and
\begin{eqnarray}
\frac{d}{dt}\left[\mathcal{P}\hat{\alpha}\left(t\right)\right] & = & \int_{0}^{t}dt^{\prime}\,\left[\mathcal{P}\mathcal{G}\left(t\right)\mathcal{G}\left(t^{\prime}\right)\mathcal{P}\right]\left[\mathcal{P}\hat{\alpha}\left(t^{\prime}\right)\right]\nonumber \\
 & = & \int_{0}^{t}dt^{\prime}\,\left[\mathcal{P}\mathcal{G}\left(t\right)\mathcal{G}\left(t^{\prime}\right)\mathcal{P}\mathcal{P}\hat{\alpha}\left(t^{\prime}\right)\right]\nonumber \\
 & = & \int_{0}^{t}dt^{\prime}\,\left[\mathcal{P}\mathcal{G}\left(t\right)\mathcal{G}\left(t^{\prime}\right)\mathcal{P}\hat{\alpha}\left(t^{\prime}\right)\right].\label{eqPalfa4}
\end{eqnarray}

Integrating Eq. (\ref{eqPalfa4}) between $t^{\prime}$ and $t$ yields
\begin{eqnarray*}
\mathcal{P}\hat{\alpha}\left(t\right)-\mathcal{P}\hat{\alpha}\left(t^{\prime}\right) & = & \int_{t^{\prime}}^{t}dt^{\prime\prime}\,\int_{0}^{t^{\prime\prime}}dt^{\prime\prime\prime}\,\left[\mathcal{P}\mathcal{G}\left(t\right)\mathcal{G}\left(t^{\prime\prime\prime}\right)\mathcal{P}\hat{\alpha}\left(t^{\prime\prime\prime}\right)\right],
\end{eqnarray*}
 which shows that the difference between $\mathcal{P}\hat{\alpha}\left(t\right)$
and $\mathcal{P}\hat{\alpha}\left(t^{\prime}\right)$ is of the second
order of magnitude in $\hat{H}_{S{\normalcolor B}}$ and, therefore,
we can write $\mathcal{P}\hat{\alpha}\left(t\right)$ instead of $\mathcal{P}\hat{\alpha}\left(t^{\prime}\right)$
in the integrand of Eq. (\ref{eqPalfa4}), obtaining an equation that
obeys the Markov approximation, without violating the Born approximation.
Thus, the master equation that we finally obtain, is written
\begin{eqnarray}
\frac{d}{dt}\left[\mathcal{P}\hat{\alpha}\left(t\right)\right] & = & \int_{0}^{t}dt^{\prime}\,\left[\mathcal{P}\mathcal{G}\left(t\right)\mathcal{G}\left(t^{\prime}\right)\mathcal{P}\hat{\alpha}\left(t\right)\right].\label{eqFINAL}
\end{eqnarray}
 In the Born approximation, to obtain the reduced density operator
of Eq. (\ref{invalfa}), we must solve Eq. (\ref{eqFINAL}) using
the initial condition
\begin{eqnarray*}
\mathrm{Tr}_{B}\left\{ \hat{\alpha}\left(0\right)\right\}  & = & \mathrm{Tr}_{B}\left\{ \hat{\rho}_{S}\left(0\right)\hat{\rho}_{B}\left(0\right)\right\} \\
 & = & \hat{\rho}_{S}\left(0\right)\mathrm{Tr}_{B}\left\{ \hat{\rho}_{B}\left(0\right)\right\} \\
 & = & \hat{\rho}_{S}\left(0\right){.}
\end{eqnarray*}

\section{Example}

As a simple example of our method, let us consider a two-level system
(a spin $\frac{1}{2}$ particle pointing along the $z$ direction)
in contact with a thermal bath of quantum harmonic oscillators:

\[
\hat{H}_{S}=\hbar\omega_{0}\hat{\sigma}_{z}
\]
 and
\[
\hat{H}_{B}=\hbar\underset{k}{\sum}\omega_{k}\hat{b}_{k}^{\dagger}\hat{b}_{k}{.}
\]
 We take the interaction to be the phase damping where, in reference
to Eq. (\ref{defHSB}),

\[
\begin{cases}
\hat{S}_{k} & =\hbar\hat{\sigma}_{z}{,}\\
\hat{B}_{k} & =g_{k}\hat{b}_{k}^{\dagger}+g_{k}^{*}\hat{b}_{k}{,}
\end{cases}
\]
 giving

\begin{eqnarray}
\hat{H}_{SB} & = & \hbar\underset{k}{\sum}\hat{\sigma}_{z}\left(g_{k}\hat{b}_{k}^{\dagger}+g_{k}^{*}\hat{b}_{k}\right){.}\label{PD}
\end{eqnarray}
 For the initial state of the thermal bath, let us consider the vacuum
state ($T=0$):

\begin{eqnarray}
\hat{\rho}_{B} & = & \left(\left|0\right\rangle \left|0\right\rangle ...\right)\otimes\left(\left\langle 0\right|\left\langle 0\right|...\right){.}\label{CIB}
\end{eqnarray}

Now we study the case of a measurement of the $z$ component, using
a single Lindblad:

\[
\hat{L}=\lambda\hat{\sigma}_{z}{,}
\]
 where $\lambda$ is a real number. Then, Eq. (\ref{defS}) gives

\begin{eqnarray*}
\mathcal{S}\hat{X} & = & -i\omega_{0}\left[\hat{\sigma}_{z},\hat{X}\right]+\lambda^{2}\left(\hat{\sigma}_{z}\hat{X}\hat{\sigma}_{z}-\hat{X}\right){.}
\end{eqnarray*}

To simplify the notation, let us define the following quantities:

\begin{equation}
\hat{R}\left(t\right)\equiv\exp\left(-\mathcal{S}t\right)\hat{\rho}_{S}\left(t\right)\label{defR}
\end{equation}
 and

\[
\mathcal{P}\hat{\alpha}\left(t\right)=\hat{R}\left(t\right)\hat{\rho}_{{\normalcolor B}}{.}
\]
 The action of ${\exp}\left({\mathcal{{S}}t}\right)$ and ${\exp}{\left(\mathcal{{B}}t\right)}$
can be calculated in the following way. For an arbitrary density operator
${\hat{X}{\left(0\right)}}{,}$ let us define

\[
\hat{X}\left(t\right)=\exp\left(\mathcal{S}t\right)\hat{X}\left(0\right){.}
\]
 Hence,

\begin{equation}
\frac{d}{dt}\hat{X}\left(t\right)=\mathcal{S}\exp\left(\mathcal{S}t\right)\hat{X}\left(0\right)=\mathcal{S}\hat{X}\left(t\right){,}\label{eqS}
\end{equation}
 that is, 
\begin{equation}
\frac{d}{dt}\hat{X}\left(t\right)=-i\omega_{0}\left[\hat{\sigma}_{z},\hat{X}\right]+\lambda^{2}\left(\hat{\sigma}_{z}\hat{X}\hat{\sigma}_{z}-\hat{X}\right){.}\label{eqXZ}
\end{equation}
 The solution of Eq. (\ref{eqXZ}) can be easily determinated \cite{key-11}:

\begin{equation}
\begin{cases}
X_{11}\left(t\right) & =X_{11}\left(0\right){,}\\
X_{12}\left(t\right) & =X_{12}\left(0\right)e^{-2\lambda^{2}t}\left[cos\left(2\omega_{0}t\right)-isen\left(2\omega_{0}t\right)\right]{.}
\end{cases}\label{solS}
\end{equation}

Analogously to the case of ${\exp}\left({\mathcal{{S}}t}\right)$,
for an arbitrary density operator ${\hat{X}{\left(0\right)}}{,}$
let us define 
\[
{\hat{X}\left(t\right)=\exp\left(\mathcal{{B}}t\right)\hat{X}\left(0\right)}{.}
\]
 From Eq. (\ref{defB}), it follows that

\[
\frac{d}{dt}\hat{X}\left(t\right)=-\frac{i}{\hbar}\left[\hat{H}_{B},\hat{X}\left(t\right)\right],
\]
 whose solution is, simply,

\begin{equation}
\hat{X}\left(t\right)=e^{-i\frac{\hat{H}_{B}}{\hbar}t}\hat{X}\left(0\right)e^{i\frac{\hat{H}_{B}}{\hbar}t}{.}\label{solB}
\end{equation}

From the above explicit actions of ${\exp}\left({\mathcal{{S}}t}\right)$
and ${\exp}{\left(\mathcal{{B}}t\right)}$ it easily follows that

\[
\begin{cases}
\exp\left(\mathcal{S}t\right)\exp\left(\mathcal{S}t'\right) & =\exp\left[\mathcal{S}\left(t+t'\right)\right]{,}\\
\exp\left(\mathcal{B}t\right)\exp\left(\mathcal{B}t'\right) & =\exp\left[\mathcal{B}\left(t+t'\right)\right]{.}
\end{cases}
\]

With Eqs. (\ref{solB}) and (\ref{solS}) we are able to solve Eq.
(\ref{eqFINAL}). Separating the system and environment terms, we
obtain, in terms of Eq. (\ref{defR}):

\begin{eqnarray*}
 & \mathcal{P}\mathcal{G}\left(t\right)\mathcal{G}\left(t'\right)\mathcal{P}\hat{\alpha}\left(t\right)=\\
 & =e^{-\mathcal{S}t}\hat{\sigma}_{z}\left\{ e^{\mathcal{S}\left(t-t'\right)}\left[\left(e^{\mathcal{S}t'}\hat{R}\left(t\right)\right)\hat{\sigma}_{z}\right]\right\} tr_{B}\left\{ e^{-\mathcal{B}t}\underset{k}{\sum}\hat{B}_{k}\left\{ e^{\mathcal{B}\left(t-t'\right)}\left[\left(e^{\mathcal{B}t'}\hat{\rho}_{B}\right)\underset{k'}{\sum}\hat{B}_{k'}\right]\right\} \right\} \otimes\hat{\rho}_{B}+\\
 & -e^{-\mathcal{S}t}\left\{ e^{\mathcal{S}\left(t-t'\right)}\left[\left(e^{\mathcal{S}t'}\hat{R}\left(t\right)\right)\hat{\sigma}_{z}\right]\right\} \hat{\sigma}_{z}tr_{B}\left\{ e^{-\mathcal{B}t}\left\{ e^{\mathcal{B}\left(t-t'\right)}\left[\left(e^{\mathcal{B}t'}\hat{\rho}_{B}\right)\underset{k}{\sum}\hat{B}_{k}\right]\right\} \underset{k'}{\sum}\hat{B}_{k'}\right\} \otimes\hat{\rho}_{B}+\\
 & -e^{-\mathcal{S}t}\hat{\sigma}_{z}\left\{ e^{\mathcal{S}\left(t-t'\right)}\left[\hat{\sigma}_{z}\left(e^{\mathcal{S}t'}\hat{R}\left(t\right)\right)\right]\right\} tr_{B}\left\{ e^{-\mathcal{B}t}\underset{k}{\sum}\hat{B}_{k}\left\{ e^{\mathcal{B}\left(t-t'\right)}\left[\underset{k'}{\sum}\hat{B}_{k'}\left(e^{\mathcal{B}t'}\hat{\rho}_{B}\right)\right]\right\} \right\} \otimes\hat{\rho}_{B}+\\
 & +e^{-\mathcal{S}t}\left\{ e^{\mathcal{S}\left(t-t'\right)}\left[\hat{\sigma}_{z}\left(e^{\mathcal{S}t'}\hat{R}\left(t\right)\right)\right]\right\} \hat{\sigma}_{z}tr_{B}\left\{ e^{-\mathcal{B}t}\left\{ e^{\mathcal{B}\left(t-t'\right)}\left[\underset{k}{\sum}\hat{B}_{k}\left(e^{\mathcal{B}t'}\hat{\rho}_{B}\right)\right]\right\} \underset{k'}{\sum}\hat{B}_{k'}\right\} \otimes\hat{\rho}_{B}{.}
\end{eqnarray*}
 Expanding the environmental superoperators using Eq. (\ref{solB})
and grouping the similar terms, we have:

\begin{eqnarray}
 & \mathcal{P}\mathcal{G}\left(t\right)\mathcal{G}\left(t'\right)\mathcal{P}\hat{\alpha}\left(t\right)=\nonumber \\
 & =\left\{ e^{-\mathcal{S}t}\hat{\sigma}_{z}\left\{ e^{\mathcal{S}\left(t-t'\right)}\left[\left(e^{\mathcal{S}t'}\hat{R}\left(t\right)\right)\hat{\sigma}_{z}\right]\right\} -e^{-\mathcal{S}t}\left\{ e^{\mathcal{S}\left(t-t'\right)}\left[\left(e^{\mathcal{S}t'}\hat{R}\left(t\right)\right)\hat{\sigma}_{z}\right]\right\} \hat{\sigma}_{z}\right\} \otimes\hat{\rho}_{B}\times\nonumber \\
 & \times tr_{B}\left\{ e^{i\frac{\hat{H}_{B}}{\hbar}t}\underset{k}{\sum}\hat{B}_{k}e^{-i\frac{\hat{H}_{B}}{\hbar}t}\hat{\rho}_{B}e^{i\frac{\hat{H}_{B}}{\hbar}t'}\underset{k'}{\sum}\hat{B}_{k'}e^{-i\frac{\hat{H}_{B}}{\hbar}t'}\right\} +\label{aux6}\\
 & +\left\{ e^{-\mathcal{S}t}\left\{ e^{\mathcal{S}\left(t-t'\right)}\left[\hat{\sigma}_{z}\left(e^{\hat{\hat{S}}t'}\hat{R}\left(t\right)\right)\right]\right\} \hat{\sigma}_{z}-e^{-\mathcal{S}t}\hat{\sigma}_{z}\left\{ e^{\mathcal{S}\left(t-t'\right)}\left[\hat{\sigma}_{z}\left(e^{\mathcal{S}t'}\hat{R}\left(t\right)\right)\right]\right\} \right\} \otimes\hat{\rho}_{B}\times\nonumber \\
 & \times tr_{B}\left\{ e^{i\frac{\hat{H}_{B}}{\hbar}t'}\underset{k}{\sum}\hat{B}_{k}e^{-i\frac{\hat{H}_{B}}{\hbar}t'}\hat{\rho}_{B}e^{i\frac{\hat{H}_{B}}{\hbar}t}\underset{k'}{\sum}\hat{B}_{k'}e^{-i\frac{\hat{H}_{B}}{\hbar}t}\right\} {.}\nonumber 
\end{eqnarray}
 Using the initial state of the environment, Eq. (\ref{CIB}), and
the interaction between the system and the environment, Eq. (\ref{PD}),
we trace out the environmental degrees of freedom and write:

\begin{eqnarray*}
 & \mathcal{P}\mathcal{G}\left(t\right)\mathcal{G}\left(t'\right)\mathcal{P}\hat{\alpha}\left(t\right)=\\
 & =\left\{ e^{-\mathcal{S}t}\hat{\sigma}_{z}\left\{ e^{\mathcal{S}\left(t-t'\right)}\left[\left(e^{\mathcal{S}t'}\hat{R}\left(t\right)\right)\hat{\sigma}_{z}\right]\right\} -e^{-\mathcal{S}t}\left\{ e^{\mathcal{S}\left(t-t'\right)}\left[\left(e^{\mathcal{S}t'}\hat{R}\left(t\right)\right)\hat{\sigma}_{z}\right]\right\} \hat{\sigma}_{z}\right\} \otimes\hat{\rho}_{B}\times\\
 & \times\underset{k}{\sum}\left|g_{k}\right|^{2}\left\{ cos\left[\omega_{k}\left(t-t'\right)\right]+isen\left[\omega_{k}\left(t-t'\right)\right]\right\} +\\
 & +\left\{ e^{-\mathcal{S}t}\left\{ e^{\mathcal{S}\left(t-t'\right)}\left[\hat{\sigma}_{z}\left(e^{\mathcal{S}t'}\hat{R}\left(t\right)\right)\right]\right\} \hat{\sigma}_{z}-e^{-\mathcal{S}t}\hat{\sigma}_{z}\left\{ e^{\mathcal{S}\left(t-t'\right)}\left[\hat{\sigma}_{z}\left(e^{\mathcal{S}t'}\hat{R}\left(t\right)\right)\right]\right\} \right\} \otimes\hat{\rho}_{B}\times\\
 & \times\underset{k}{\sum}\left|g_{k}\right|^{2}\left\{ cos\left[\omega_{k}\left(t-t'\right)\right]-isen\left[\omega_{k}\left(t-t'\right)\right]\right\} 
\end{eqnarray*}

Now we rewrite the system superoperators using Eq. (\ref{solS}).
Thus, in terms of $\hat{R}\left(t\right){,}$ we have
\begin{eqnarray*}
 & \mathcal{P}\mathcal{G}\left(t\right)\mathcal{G}\left(t'\right)\mathcal{P}\hat{\alpha}\left(t\right)=\\
= & -4\left(\begin{array}{cc}
0 & R_{12}\\
R_{21} & 0
\end{array}\right)\underset{k}{\sum}\left|g_{k}\right|^{2}cos\left[\omega_{k}\left(t-t'\right)\right]{,}
\end{eqnarray*}
 which, in accordance with Eq. (\ref{eqFINAL}), gives

\begin{equation}
\frac{d}{dt}\left(\begin{array}{cc}
R_{11} & R_{12}\\
R_{21} & R_{22}
\end{array}\right)=-4\left(\begin{array}{cc}
0 & R_{12}\\
R_{21} & 0
\end{array}\right)\int_{0}^{t}dt'\underset{k}{\sum}\left|g_{k}\right|^{2}cos\left[\omega_{k}\left(t-t'\right)\right]{.}\label{eqmat}
\end{equation}

Now we make the continuum transformation \cite{key-2}. Let us define
the density of states as

\[
J\left(\omega\right)=\underset{k}{\sum}\left|g_{k}\right|^{2}\delta\left(\omega-\omega_{k}\right){,}
\]
 which allows us to rewrite Eq. (\ref{eqmat}) as

\[
\frac{d}{dt}\left(\begin{array}{cc}
R_{11} & R_{12}\\
R_{21} & R_{22}
\end{array}\right)=-4\left(\begin{array}{cc}
0 & R_{12}\\
R_{21} & 0
\end{array}\right)\int_{0}^{t}dt'\int_{0}^{\infty}d\omega J\left(\omega\right)cos\left[\omega\left(t-t'\right)\right]
\]
 and, with the change of variable $\tau=t-t'{,}$ we obtain

\begin{equation}
\frac{d}{dt}\left(\begin{array}{cc}
R_{11} & R_{12}\\
R_{21} & R_{22}
\end{array}\right)=-4\left(\begin{array}{cc}
0 & R_{12}\\
R_{21} & 0
\end{array}\right)\int_{0}^{t}d\tau\int_{0}^{\infty}d\omega J\left(\omega\right)cos\left(\omega\tau\right){.}\label{eqtau}
\end{equation}

To obtain the solution for the diagonal terms of $\hat{R}\left(t\right)$,
we do not need further consideration:

\[
\begin{cases}
R_{11}\left(t\right)= & R_{11}\left(0\right){,}\\
R_{22}\left(t\right)= & R_{22}\left(0\right){.}
\end{cases}
\]
 However, for the non-diagonal terms, it is necessary to specify $J\left(\omega\right){.}$
Let us use the Ohmic density of states:

\[
J\left(\omega\right)=\eta\omega e^{-\frac{\omega}{\Omega}}{,}
\]
 where $\eta$ and $\Omega$ are real and positive constants. Hence,
Eq. (\ref{eqtau}) becomes 
\[
\frac{d}{dt}\left(\begin{array}{cc}
R_{11} & R_{12}\\
R_{21} & R_{22}
\end{array}\right)=-4\eta\left(\begin{array}{cc}
0 & R_{12}\\
R_{21} & 0
\end{array}\right)\int_{0}^{t}d\tau\int_{0}^{\infty}d\omega\omega e^{-\frac{\omega}{\Omega}}cos\left(\omega\tau\right){.}
\]
 The double integral gives

\[
\int_{0}^{t}d\tau\int_{0}^{\infty}d\omega\omega e^{-\frac{\omega}{\Omega}}cos\left(\omega\tau\right)=\frac{\Omega^{2}t}{1+\left(\Omega t\right)^{2}}{,}
\]
 i.e.,

\[
\begin{cases}
\frac{d}{dt}R_{12}= & -4\eta\frac{\Omega^{2}t}{1+\left(\Omega t\right)^{2}}R_{12}{,}\\
\frac{d}{dt}R_{21}= & -4\eta\frac{\Omega^{2}t}{1+\left(\Omega t\right)^{2}}R_{21}{.}
\end{cases}
\]
 The solution for this system is

\[
\begin{cases}
R_{12}\left(t\right)= & \frac{R_{12}\left(0\right)}{\left[1+\left(\Omega t\right)^{2}\right]^{2\eta}}{,}\\
R_{21}\left(t\right)= & \frac{R_{21}\left(0\right)}{\left[1+\left(\Omega t\right)^{2}\right]^{2\eta}}{.}
\end{cases}
\]
 At last, now we calculate the density-operator elements by inverting
Eq. (\ref{defR}) and we obtain

\begin{equation}
\begin{cases}
\rho_{11}\left(t\right)= & \rho_{11}\left(0\right){,}\\
\rho_{12}\left(t\right)= & \rho_{12}\left(0\right)\frac{e^{-2\lambda^{2}t}}{\left[1+\left(\Omega t\right)^{2}\right]^{2\eta}}\left[cos\left(2\omega_{0}t\right)-isen\left(2\omega_{0}t\right)\right]{,}
\end{cases}\label{aux7}
\end{equation}
 remembering that $\rho_{22}\left(t\right)=1-\rho_{11}\left(t\right)$
and $\rho_{21}\left(t\right)=\rho_{12}^{*}\left(t\right){.}$

\section{Conclusion}

In summary, here we present a new method to describe the dynamics
of measurements that occur during a finite time interval, while the
system being measured interacts with the rest of the universe and,
due to the consequent environmentally-induced noise, undergoes decoherence.
We use a Lindbladian description of the measuring apparatus, whose
interaction with the system we assume as Markovian. To treat the noise
introduced by the fact that, during the finite-duration measurement,
the system is perturbed by the environment, we use a Redfield approach
to the dynamical description of the interaction between the system
and its environment, assumed non-Markovian. The resulting unprecedented
hybrid description is shown to be capable of substantially simplifying
the tracing procedure, which is usually very complicated due to the
non-commutativity of several terms in the total Hamiltonian, comprising
the system under scrutiny, its environment, and the observables to
be measured.

The superoperators defined in Sec. 2 introduce simplifications of
the calculations leading to Eq. (\ref{eqFINAL}), that is compact
and can be solved in terms of the unperturbed solutions (in Sec. 5,
for example, we used Eq. (\ref{solS})). Moreover, for the Born-Markov
approximation, regardless of the model chosen for the environment,
the reduction of the density operator becomes evident, as can be verified
in Eq. (\ref{aux6}). The simple phase-damping-interaction example
of Sec. 5 (see Eq. (\ref{PD})) at zero temperature already provides
an important and expected result in the quantum mechanics of open
systems: the intensification of the environmentally-induced decoherence,
as indicated in the denominator of the second of Eqs. (\ref{aux7}),
evidencing the power and convenience of the present approach.

\section*{Acknowledgement}

The authors wish to thank the Coordenação de Aperfeiçoamento de Pessoal
de Nível Superior (CAPES) and the Conselho Nacional de Desenvolvimento
Científico e Tecnológico (CNPq), Brazil. This work has also been supported
by Fundação de Amparo à Pesquisa do Estado de São Paulo, Brazil, project
number 05/04105-5 and the Millennium Institute for Quantum Information
- Conselho Nacional de Desenvolvimento Científico e Tecnológico, Brazil.

\end{document}